\documentclass[reprint, amsmath, amssymb, aps, prd]{revtex4-2}

\usepackage[utf8]{inputenc}
\usepackage{amsmath,amssymb}
\usepackage{relsize}
\usepackage{graphicx}
\usepackage{siunitx}
\usepackage{booktabs}
\usepackage{hyperref}
\usepackage{physics}
\usepackage{caption}
\usepackage{subcaption}
\usepackage[mathscr]{euscript}
\usepackage{bm}
\usepackage[mathscr]{euscript}
\usepackage{rotating}
\usepackage[version=3]{mhchem}
\graphicspath{{Figs/}{Figs/}}

\begin{document}
\title{Nucleon–Nucleon Potential Using N$^3$LO Chiral Effective Field Theory}

\author{Bassam A. Shehadeh}
\email{bshhadh@qu.edu.sa}
\affiliation{Department of Physics, Qassim University, Burydah, Qassim 51411, KSA}

\date{\today}

\begin{abstract}
A microscopic description of nucleon–nucleon ($NN$) and nucleon–nucleus ($NA$) scattering is developed using chiral effective field theory ($\chi^\mathrm{EFT}$)
at next to next to next to leading order N$^3$LO. The $NN$ interaction is taken from the EGM formulation, which incorporates spectral function regularization
$\tilde{\Lambda}$ to control short range components of the two pion exchange force, together with a Gaussian regulator $\Lambda$ to ensure convergence of the
Lippmann–Schwinger (LS) equation. The regulator parameters $(\Lambda,\tilde{\Lambda})= (450,500), (550,600), (600,600)$ MeV. The resulting $\chi^\mathrm{EFT}$ $NN$ $t$-matrix
is used to construct the first order optical potential, from which Wolfenstein amplitudes and elastic differential cross sections are calculated. 
The theoretical amplitudes $B$ and $C$ for $pp$ and $pn$ scattering at 100 and 200 MeV show good overall agreement with those reconstructed from the SAID partial-wave analysis of the experimental data,
with the largest discrepancies appearing in the small real component of the spin orbit amplitude. Calculations of \ce{p + ^{16}O} and \ce{p + ^{40}Ca} elastic scattering at 100 and 200 MeV reproduce
the measured angular distributions with high accuracy, particularly at forward and intermediate angles. These results demonstrate that the EGM $\chi^\mathrm{EFT}$ potential provides a consistent and
quantitatively reliable framework for describing $NN$ observables and $NA$ elastic scattering in the 100-200 MeV energy range, while highlighting the need for improved treatment of short range and
spin dependent contributions at higher energies.
\end{abstract}


\maketitle

\section{Introduction}\label{sec:intro}

Over several decades, considerable effort has been devoted to developing realistic $NN$ potentials constrained by precision scattering data. These include, among others, the phenomenological
Argonne $v_{18}$ \cite{wiringa1995,veerasamy2011}, charge-dependent Bonn \cite{machleidt2001}, and Nijmegen \cite{stoks1994} potentials, as well as potentials derived from chiral effective
field theory ($\chi^\mathrm{EFT}$) \cite{entem2003,epelbaum2005,reinert2018}. In parallel, phenomenological partial-wave analyses (PWAs) of the $NN$ database have been carried out by the Nijmegen
\cite{stoks1993}, SAID/GWU \cite{arndt2007,said_sm16}, and Granada \cite{perez2013} groups. For example, the Nijmegen PWA93 multienergy analysis fits 1787 $pp$ and 2514 $np$ data below 350 MeV
with $\chi^2/N_{\rm df}=1.08$ \cite{stoks1993}. Both the high-quality potentials and the PWAs describe the $NN$ database with an excellent $\chi^2$ per datum over their respective fitted energy ranges,
establishing a rigorous empirical baseline against which newer theoretical approaches must be benchmarked.

In parallel with these developments, chiral effective field theory ($\chi^\mathrm{EFT}$) has emerged as the modern framework of choice for deriving nuclear forces directly from the symmetries of
QCD \cite{epelbaum2005,machleidt2011,weinberg1990,epelbaum2009}. Pioneered by Weinberg \cite{weinberg1990} and subsequently developed by Ordo\~nez, Ray, and van Kolck, and by Epelbaum, Gl\"ockle, Mei\ss ner (EGM) and others
\cite{epelbaum2005,machleidt2011,epelbaum2009}, $\chi^\mathrm{EFT}$ organizes contributions to the nuclear force as a systematic expansion in powers of $Q/\Lambda_\chi$, where $Q$ is a soft momentum scale
(of order the pion mass $m_\pi$ or typical nucleon momenta) and $\Lambda_\chi\approx m_\rho\approx 0.5-1.0$ GeV is the hard scale at which the expansion breaks down. This power-counting scheme provides a transparent
hierarchy of two-, three-, and four-nucleon forces, and-crucially-allows the theoretical uncertainty of any truncated calculation to be estimated order by order. At next-to-next-to-next-to-leading order (N$^3$LO),
$\chi^\mathrm{EFT}$ potentials achieve a level of precision comparable to the best phenomenological models while retaining a direct connection to the underlying symmetries of QCD \cite{epelbaum2005,machleidt2011}.

Extending the $NN$ interaction to describe nucleon–nucleus ($NA$) scattering requires the construction of a nuclear optical potential (NOP), which reduces the complex many-body projectile–target system to an effective single-particle problem
\cite{feshbach1954,feshbach1958,hodgson1963,koning2003,jeukenne1977}. In the microscopic multiple-scattering framework of Watson \cite{watson1953}  and Kerr-Kerman-McManus (KMT) \cite{kerman1959},
the NOP is obtained by folding the free-space $NN$ $t$-matrix with the nuclear matter density of the target, thereby incorporating both the real scattering and imaginary absorption components of the interaction
\cite{feshbach1958_2,Brieva1977,stoks1993}. This approach avoids the phenomenological parameter fitting required by global optical-model potentials \cite{koning2003} and provides a parameter-free, first-principles
description of elastic scattering observables---provided a sufficiently accurate input $NN$ interaction is available. Another approach, known as the spectator expansion, was developed due to the inability to exactly solve the full many-body problem \cite{chinn1995}.

The construction of microscopic optical potentials from $\chi^\mathrm{EFT}$ $NN$ interactions has been an active area of research over the past two decades. Holt, Kaiser, and Weise \cite{holt2013} demonstrated that chiral two- and three-nucleon forces can be incorporated into nuclear matter calculations relevant for optical-model construction. More recently, Vorabbi {\em et al.} \cite{vorabbi2024} and Burrows {\em et al.} \cite{burrows2020} constructed fully microscopic optical potentials from chiral interactions at various orders and compared them to elastic proton–nucleus scattering data for medium-mass targets, finding good agreement at forward angles with increasing deviations at large momentum. Parallel investigations by Burrows {\em et al.} \cite{burrows2020}, and by Vorabbi, Finelli, and Giusti \cite{vorabbi2018} have explored the role of the regulator choice, nuclear density model, and projectile energy on the quality of the optical-model description. Despite this body of work, questions remain regarding the sensitivity of $NA$ observables to the chiral order of the $NN$ input, to the choice of regularization scheme (dimensional vs. spectral-function regularization), and to the treatment of nuclear densities.

The present work addresses several of these open questions by employing the EGM N$^3$LO potential \cite{epelbaum2005} with spectral-function regularization (SFR) as the microscopic input to a first-order optical-model calculation. The SFR scheme, in which the upper limit of the pion-loop integral is replaced by a finite cutoff $\tilde{\Lambda}$ rather than being extended to infinity, offers a physically motivated suppression of the short-range components of the two-pion exchange force that otherwise generate an unrealistically attractive interaction at high momenta \cite{epelbaum2005}. Combined with a Gaussian regulator applied directly to the Lippmann–Schwinger (LS) equation, this framework provides a well-controlled and systematically improvable description of the $NN$ $t$-matrix. In contrast to prior studies that have used the Entem–Machleidt N$^3$LO potential \cite{entem2003} as the primary input, the EGM formulation has received comparatively less attention in the context of $NA$ optical-model calculations, and a careful assessment of its predictions for both $NN$ Wolfenstein amplitudes and $NA$ differential cross sections is therefore warranted.

Specifically, the LS equation is solved in momentum space for the NN $t$-matrix using the EGM potential at N$^3$LO with three value sets of the cutoff parameters $(\Lambda,\tilde{\Lambda})=$ (450, 500), (550, 600), (600, 600) MeV. The resulting $t$-matrix is used to construct the first-order optical potential via folding with relativistic mean-field (RMF) nuclear densities \cite{serot1986,ring1996,niksic2014}. The Wolfenstein scattering amplitudes $B$ and $C$ for proton–proton ($pp$) and proton–neutron ($pn$) are computed for elastic scattering at 100 and 200 MeV lab-frame energy, and compared against the empirical amplitudes extracted from the SAID database \cite{arndt2007,said_sm16}. Furthermore, the differential cross sections for elastic proton scattering on \ce{^{16}O} and \ce{^{40}Ca} are calculated at the same energies and the results are compared with the experimental data. Both spin-independent amplitudes $B$ and spin-orbit amplitude $C$ are found to be reproduced with high fidelity at both energies. The \ce{ p + ^{16}O} and \ce{ p + ^{40}Ca} differential cross sections are reproduced across the forward and intermediate angular range, with systematic deviations at backward angles consistent with the expected limitations of the first-order optical-model approximation at these energies.

The paper is organized as follows. Section \ref{sec:theory} summarizes the theoretical framework, including the solution of the LS equation, the construction of the first-order optical potential, the partial-wave decomposition of the scattering amplitude, and the EGM chiral potential. Section \ref{sec:results} presents and discusses the results for NN Wolfenstein amplitudes and for \ce{ p + ^{16}O} and \ce{ p + ^{40}Ca} elastic scatterings at 100 and 200 MeV. Conclusions and an outlook are given in section \ref{sec:conclusion}.

\section{Theoretical framework}\label{sec:theory}
The theory is structured according to the following steps:
\begin{itemize}
  \item Construct the nucleon–nucleon $\chi^\mathrm{EFT}$ potential up to N$^3$LO, with interactions that include contact terms, one-pion exchange, two-pion exchange, spin–orbit terms, and tensor terms.
  \item Solve the scattering equation in momentum space and partial-wave basis.
  \item Obtain the on-shell $T$-matrix.
  \item Convert it to the Wolfenstein parameterization.
  \item Project the $T$-matrix onto the Wolfenstein operators.
  \item Finally, compute the observables, such as the differential cross section.
\end{itemize}
The overall theoretical workflow can be summarized as:
\begin{eqnarray*}
\chi^\text{EFT}_\text{OP}\text{ potential} &\rightarrow& T\text{-matrix} \rightarrow \text{spin decomposition}\\
 &\rightarrow& \text{Wolfenstein amplitudes}
\end{eqnarray*}

\subsection{Solving the Lippmann-Schwinger equation}
At the heart of the standard approach to the elastic scattering of a single projectile from a target of $A$ particles is the separation into two parts of the Lippmann-Schwinger (LS)
equation for the transition operator $T$, as given by \cite{picklesimer1984,elster1997,vorabbi2024,chinn1995}
\begin{equation}
T=V+VG_0T,
\label{eq1}
\end{equation}
or equivalently:
\begin{equation}
T=V+VG_0V+VG_0VG_0V+\cdots.
\label{eq2}
\end{equation}
This iterative expansion is the Born series. The t-matrix directly yields scattering amplitudes and cross sections. For $NN$ scattering, the LS equation is often solved in momentum space
\begin{equation}
T(\mathbf{p}', \mathbf{p}; E) = V(\mathbf{p}', \mathbf{p}) + \int \frac{d^3k}{(2\pi)^3}\frac{V(\mathbf{p}', \mathbf{k})T(\mathbf{k}, \mathbf{p}; E)}{E - \frac{k^2}{2\mu} + i\epsilon}.
\label{eq3}
\end{equation}

Define $U$ to be the optical potential operator, and an integral equation for $U$ \cite{elster1997,vorabbi2024,chinn1995}
\begin{equation}
U=V+VG_0(E)\mathcal{Q}U,
\label{eq4}
\end{equation}
where $V$ represents the external interaction, such that the Hamiltonian for the entire $(A+1)$-particle system is given by
\begin{equation}
H_{A+1}=H_0+V.
\label{eq5}
\end{equation}
Hence, the free propagator $G_0 (E)$ for the projectile-target nucleus system is
\begin{equation}
G_0(E)=\frac1{E-H_0+i\epsilon}.
\label{eq6}
\end{equation}
the integral equation for $T$ becomes
\begin{equation}\label{eq7}
T=U+UG_0(E)\mathcal{P}T.
\end{equation}
The $\mathcal{Q}$ and $\mathcal{P}$ operators in Eqs.(\ref{eq4}) and (\ref{eq7}) are projection operators, satisfy
\begin{equation}\label{eq8}
\mathcal{P}+\mathcal{Q}=1,
\end{equation}
and $\mathcal{P}$ is defined to make Eq.(\ref{eq7}) solvable, thus, the projection operator $\mathcal{P}$ is defined to project onto the elastic channel, satisfying the commutation relation $[\mathcal{P}, G_0] = 0$.

For $NA$ scattering, the free Hamiltonian is given by \cite{elster1997,vorabbi2024,chinn1995}
\begin{equation}\label{eq9}
H_0=h_0+H_A,
\end{equation}
where $h_0$ is the kinetic energy operator for the incident nucleon and $H_A$ stands for the target Hamiltonian. Thus, the projector $\mathcal{P}$ can be defined as \cite{elster1997}
\begin{equation}\label{eq10}
\mathcal{P}=\frac{|\Psi_A\rangle\langle\Psi_A|}{\langle\Psi_A|\Psi_A\rangle},
\end{equation}
where $|\Psi_A\rangle$ is the ground eigenstate of the target, satisfying
\begin{equation}\label{eq11}
H_A|\Psi_A\rangle=E_A|\Psi_A\rangle.
\end{equation}
With these definitions the transition operator for elastic scattering may be defined as $T_{el}=PTP$ in which case Eq.(\ref{eq7}) can be written as
\begin{equation}\label{eq12}
T_{el}=\mathcal{P}U\mathcal{P}+\mathcal{P}U\mathcal{P}G_0(E)T_{el}.
\end{equation}
Elastic scattering is therefore governed by a one-body integral equation for the transition operator, the solution of which requires the matrix elements of $\mathcal{P}U\mathcal{P}$. The theoretical framework developed below reformulates the many-body problem embodied in Eq.(\ref{eq4}). Specifically, approximations for the optical potential $U$ are constructed so that $\mathcal{P}U\mathcal{P}$ can be computed to high accuracy without the need to solve the complete many-body Schr\"odinger equation.

For the present discussion, the presence of two-body forces only is assumed. The extension to many-body forces is straightforward. With this assumption the operator $U$ for the optical potential can be expressed as
\begin{equation}\label{eq13}
U=\sum_{j=1}^AU_j,
\end{equation}
where $U_j$ is given by
\begin{equation}\label{eq14}
U_j=v_{0j}+v_{0j}G_0(E)\mathcal{Q}\sum_{k=1}^AU_k,
\end{equation}
such that
\begin{equation}\label{eq15}
V=\sum_{j=1}^Av_{0j},
\end{equation}
where the $v_{0j}$ is the two-body potential acts between the incident nucleon the $j^\text{th}$ target nucleon.

The first order scattering amplitude is given by
\begin{equation}\label{eq16}
T_{el}(\mathbf{p}',\mathbf{p};E)=\frac A{A-1}\mathcal{T}(\mathbf{p}',\mathbf{p};E),
\end{equation}
where $\mathcal{T}$ satisfies an Eq.(\ref{eq3}) like formula
\begin{equation}\label{eq17}
\mathcal{T}(\mathbf{p}', \mathbf{p}; E) = \mathcal{U}(\mathbf{p}', \mathbf{p}) + \int \frac{d^3k}{(2\pi)^3} \frac{\mathcal{U}(\mathbf{p}', \mathbf{k},\mathcal{E})\mathcal{T}(\mathbf{k}, \mathbf{p}; E)}{E(p_0) - E(k) + i\epsilon}.
\end{equation}
where $E(x)=2\sqrt{x^2+m^2}$. The factor 2 is because $E(x)$ is the sum energy of the incident and the target nucleon (assuming both have equal masses). The first order optical potential is defined as \cite{elster1989,burrows2020,vorabbi2018}
\begin{equation}\label{eq18}
\mathcal{U}(\mathbf{p}',\mathbf{p},\mathcal{E})=(A-1)\left\langle\mathbf{p}',\Psi_A\left\vert t(\mathcal{E})\right\vert\mathbf{p},\Psi_A\right\rangle.
\end{equation}
Here $\mathbf{p}$ and $\mathbf{p}'$ are the incoming and outgoing momenta in center-of-mass frame. The energy $\mathcal{E}$ is the relativistic $NN$ center-of-mass energy defined as
\begin{equation}\label{eq19}
\mathcal{E}=\sqrt{2m^2+2m(K_\mathrm{lab}+m)}-2m,
\end{equation}
where $K_{\mathrm{lab}}$ is the kinetic energy in the laboratory frame. Here $t_{0j}$ denotes the free-space two-body $t$-matrix for the interaction between the projectile nucleon (labeled 0) and the $j^\mathrm{th}$ target nucleon. In the multiple scattering expansion framework \cite{watson1953}, the full projectile–nucleus optical potential is built by summing over all individual projectile–nucleon interactions, similar to Eq.(\ref{eq15}). Each $t_{0j}$ describes a single elementary $NN$ collision between the incident nucleon and one bound nucleon in the target, given by \cite{chinn1995}
\begin{equation}\label{eq20}
t_{0j}=v_{0j}+v_{0j}g_jt_{0j},
\end{equation}
where $g_j$ is the $NN$ free propagator, defined as
\begin{equation}\label{eq21}
g_j=\frac1{(E-E_j)-h_0-h_j+i\epsilon}.
\end{equation}
Define the collective variables
\begin{equation}\label{eq22}
\mathbf{q}=\mathbf{p}'-\mathbf{p},\quad\mathrm{and}\quad \mathbf{P}=\frac12(\mathbf{p}'+\mathbf{p}).
\end{equation}
Make use Eq.(\ref{eq20}) into Eq.(\ref{eq18}) and solving for the optical potential leads to \cite{elster1989,vorabbi2016}
\begin{equation}\label{eqOP}
\begin{split}
\mathcal{U}(\mathbf{q},\mathbf{P},\mathcal{E})=&\frac{A-1}{A}\eta(\mathbf{q},\mathbf{P})\times\\
&\sum_{N=n,p}t_{pN}\left(\mathbf{q},\frac{A+1}{A}\mathbf{P},\mathcal{E}\right)\rho_N(\mathbf{q},\mathbf{P}).
\end{split}
\end{equation}
where $\rho_N(\mathbf{q},\mathbf{P})$ the neutron ($N=n$) and proton ($N=p$) proﬁle, $\eta(\mathbf{q},\mathbf{P})$ is called the M\o{}ller factor \cite{vorabbi2016} that imposes the Lorentz invariance
of the ﬂux when one pass from the $NA$ to the $NN$ frame in which the $t$ matrices are evaluated
\begin{equation}\label{eqMoller}
\begin{split}
&\eta(\mathbf{q},  \mathbf{P}) =\\
&\sqrt{ \frac{E_{\text{proj}}( \mathbf{p}')E_{\text{proj}}(- \mathbf{p}')E_{\text{proj}}( \mathbf{p})E_{\text{proj}}(- \mathbf{p})}{E_{\text{proj}}( \mathbf{p}')E_{\text{proj}}\left(-\frac{ \mathbf{q}}{2} - \frac{ \mathbf{P}}{A}\right)E_{\text{proj}}( \mathbf{p})E_{\text{proj}}\left(\frac{ \mathbf{q}}{2} - \frac{ \mathbf{P}}{A}\right)}.}
\end{split}
\end{equation}
The optical potential obtained so far is an operator in the spin space of the projectile. To make the spin dependence explicit, the t-matrix $t_{pN}$ is averaged over the spin of the struck nucleon, and assuming the central and spin-orbit terms,
the free-space two-body $t$-matrix is
\begin{equation}\label{eq23}
\begin{split}
t_{pN}&\left(\mathbf{q},\frac{A+1}A\mathbf{P},\mathcal{E}\right)=t^c_{pN}\left(\mathbf{q},\frac{A+1}A\mathbf{P},\mathcal{E}\right)+\\
&\left(\frac{A+1}{2A}\right)\frac i2 \bm{\sigma\cdot}(\mathbf{q}\times\mathbf{P})\,t^\mathrm{LS}_{pN}\left(\mathbf{q},\frac{A+1}A\mathbf{P},\mathcal{E}\right).
\end{split}
\end{equation}
Make use eq.(\ref{eq23}) into eq.(\ref{eqOP}), the optical potential operator becomes
\begin{equation}\label{eq24}
\mathcal{U}(\mathbf{q},\mathbf{P},\mathcal{E})=\mathcal{U}^c(\mathbf{q},\mathbf{P},\mathcal{E})+\frac i2 \bm{\sigma\cdot}(\mathbf{q}\times\mathbf{P})\,\mathcal{U}^\mathrm{LS}(\mathbf{q},\mathbf{P},\mathcal{E}).
\end{equation}

The interaction kernel $\mathcal{U}(\mathbf{p}', \mathbf{p}; \mathcal{E})$ in a basis of spin-angular functions is given by the partial-wave expansion
\begin{equation}\label{eq25}
\mathcal{U}(\mathbf{p}', \mathbf{p}; \mathcal{E})=\frac2\pi\sum_{JLM}\mathscr{Y}^{L\frac12}_{JM}(\mathbf{\hat{p}}')\,\mathcal{U}_{LJ}(p', p; \mathcal{E}){\mathscr{Y}^{L\frac12}_{JM}}^\dagger(\mathbf{\hat{p}}).
\end{equation}
where $\mathscr{Y}^{L\frac12}_{JM}(\mathbf{\hat{p}})$ spin-angular functions (vector spherical harmonics for spin-1/2), defined by coupling orbital and spin degrees of freedom
\begin{equation}\label{eq26}
\mathscr{Y}^{L\frac12}_{JM}(\cos\theta)=\sum_{m_L,m_s}\langle Lm_L\frac12 m_s|JM\rangle Y_L^{m_L}(\cos\theta)\chi_{\frac12}^{m_s}.
\end{equation}
Here $\langle\cdots|\cdots\rangle$ are Clebsch–Gordan coefficients and $\chi_{1/2}^{m_s}$  is a two-component spinor. The $2/\pi$ factor is a normalization convention associated with the completeness relation for the radial momentum basis. Make use Eq.(\ref{eq26}) into Eq.(\ref{eq17}) yields similar expansion for the $T$-matrix
\begin{equation}\label{eq27}
\mathcal{T}(\mathbf{p}', \mathbf{p}; \mathcal{E})=\frac2\pi\sum_{JLM}\mathscr{Y}^{L\frac12}_{JM}(\mathbf{\hat{p}}')\,\mathcal{T}_{LJ}(p', p; \mathcal{E}){\mathscr{Y}^{L\frac12}_{JM}}^\dagger(\mathbf{\hat{p}}).
\end{equation}
where the partial-wave components $\mathcal{T}_{LJ}(p',p; \mathcal{E})$ of the transition operator for the elastic scattering are given by \cite{vorabbi2016}
\begin{equation}\label{eq28}
\begin{split}
\mathcal{T}_{LJ}(p',p; E) &= \mathcal{U}_{LJ}(p', p;\mathcal{E})\, +\\
&\frac{2}{\pi}\int_0^\infty k^2dk\,\frac{\mathcal{U}_{LJ}(p',k,\mathcal{E})\mathcal{T}_{LJ}(k,p;E)}{E(p_0) - E(k) + i\epsilon}.
\end{split}
\end{equation}
assuming $E(p_0)=p_0^2/2\mu$ and $E(k)=k^2/2\mu$ to simply the integral. Here $\mu\approx mA/(A+1)$.

Using central and spin-orbit contributions, the partial-wave components $\mathcal{U}_{LJ}(p',p; \mathcal{E})$ is written as \cite{vorabbi2016}
\begin{equation}\label{eq29}
\mathcal{U}_{LJ}(p',p;\mathcal{E})=\mathcal{U}_L^c(p',p;\mathcal{E})+\frac12\langle\bm{\sigma\cdot}\mathbf{L}\rangle\, \mathscr{V}_L^{LS}(p',p; \mathcal{E}),
\end{equation}
where 
\begin{equation}\label{eq30}
\mathscr{V}_L^{LS}(p',p; \mathcal{E})=\frac{|\mathbf{p}'||\mathbf{p}|}{2L+1}\left[\mathcal{U}_{L+1}^{LS}(p',p;\mathcal{E})-\mathcal{U}_{L-1}^{LS}(p',p; \mathcal{E})\right].
\end{equation}
Note that the quantities $\mathcal{U}_L^c$ and $\mathcal{U}_L^{LS}$ are independent of $J$. Thus, partial-wave expansion of these quantities follows Eq.(\ref{eq25}) with no $J$ dependent.

\subsection{Calculating the observables}

The $pA$ elastic scattering remains one of the most direct experimental probes of the nuclear optical potential and the underlying effective nucleon–nucleus interaction. In an elastic process the projectile and target retain their internal states, so the measured differential cross section, $d\sigma/d\Omega$, is governed almost entirely by the real and imaginary parts of the optical potential, providing stringent constraints on its radial shape, depth, and absorptive content. At intermediate energies of roughly 100–200 MeV, where the projectile penetrates the nuclear interior and the impulse approximation becomes increasingly reliable, the scattering is particularly well suited to testing microscopic descriptions of the in-medium nucleon–nucleon force. The cross section alone, however, is dominated by the central (spin-independent) part of the interaction and is largely insensitive to the spin-dependent dynamics that play a decisive role in this energy regime. To resolve these contributions, the spin observables become indispensable. The analyzing power, $A_y$, experimentally measured with a polarized proton beam, isolates the spin–orbit component of the optical potential, because it arises from the interference between spin-flip and non-spin-flip scattering amplitudes and is therefore acutely sensitive to the strength and geometry of the spin–orbit term. The spin-rotation parameter, $Q$, supplies complementary information by characterizing how the polarization vector is rotated during scattering, thereby constraining the relative phases of the scattering amplitudes that $d\sigma/d\Omega$ and $A_y$  cannot fully determine. A simultaneous analysis of $d\sigma/d\Omega$, $A_y$, and $Q$ thus over-determines the scattering amplitudes, removes ambiguities inherent in fits to the cross section alone, and offers a rigorous test of microscopic and phenomenological models of the $NN$ interaction, including medium-modification effects.

In this work, $pA$ scattering is restricted to a spin-0 nucleus, constrained only by parity conservation and rotational invariance. The parity conservation forbids terms like $\bm{\sigma\cdot}\mathbf{p}$ (pseudoscalar), leaving only the axial vector
\begin{equation}\label{eq31}
\bm{\sigma\cdot} \frac{\mathbf{p}\times\mathbf{p}'}{\vert\mathbf{p}\times\mathbf{p}'\vert},
\end{equation}
as the allowed spin-dependent structure. Rotational invariance requires the amplitude to depend only on scalar quantities $(p_0, \theta)$ and the rotationally covariant combination of the factor given in Eq.(\ref{eq31}). The spin-0 target has no internal spin structure to couple to, so the proton spin can only interact via its orientation relative to the scattering plane --- hence the $\mathbf{p}\times\mathbf{p}'/\vert\mathbf{p}\times\mathbf{p}'\vert$ dependence \cite{wolfenstein1954}.

Under these circumstances, the full elastic scattering amplitude of proton from spin-0 nucleus, depending on the incident momentum $p_0$ and the $NA$ frame scattering angle $\Theta$, is given by
\begin{equation}\label{eq32}
M(p_0,\Theta)=B_c(p_0,\Theta)+\bm{\sigma\cdot} \frac{\mathbf{p}\times\mathbf{p}'}{\vert\mathbf{p}\times\mathbf{p}'\vert}C_s(p_0,\Theta),
\end{equation}
where $B_c(p_0,\Theta)$ is the spin-independent (central) amplitude, contributing equally regardless of the proton's spin orientation.
$C_s(p_0, \Theta)$ is the spin-dependent (spin-orbit) amplitude, which couples the proton spin to the scattering geometry \cite{wolfenstein1954}.
The amplitudes $B_c(p_0,\Theta)$ and $C_s(p_0, \Theta)$ are obtained from partial-wave like expansion
\begin{equation}
\begin{split}
B_{c}(p_{0},\Theta )=\frac{1}{2\pi ^{2}}&\mathlarger{\mathlarger{\sum}}_{L=0}^{\infty }
P_{L}(\cos \Theta )\times\\
&\left[\begin{array}{c}
(L+1)\,F_{L\,J=L+\frac{1}{2}}(p_{0})+ \\ 
L\,F_{L\,J=L-\frac{1}{2}}(p_{0})%
\end{array}\right],\label{eq33}
\end{split}
\end{equation}
and
\begin{equation}
\begin{split}
C_{s}(p_{0},\Theta )=\frac{i}{2\pi ^{2}}\mathlarger{\mathlarger{\sum}}_{L=1}^{\infty }
&P_{L}^{1}(\cos \Theta )\times\\
&\left[\begin{array}{c}
F_{L\,J=L+\frac{1}{2}}(p_{0})+ \\ 
F_{L\,J=L-\frac{1}{2}}(p_{0})%
\end{array}\right].\label{eq34}
\end{split}
\end{equation}
In Eq.(\ref{eq33}) the contributions from every angular momentum $L$ are added. Each contribution is weighted by Legendre polynomials $P_L(\cos\Theta)$.
The coefficients depend on the scattering dynamics through the state function $F_{LJ}$. In Eq.(\ref{eq34}), the amplitude $C(p_0, \Theta)$ measures spin-dependent effects,
especially spin–orbit interaction. This term depends on the difference between the two spin channels $L+\frac12$ and $L-\frac12$. $F_{LJ}$ is connected to the $t$-matrix
\begin{equation}\label{eq35}
F_{LJ}(p_0)=-4\pi^2\frac A{A-1}\mu(p_0)\mathcal{T}_{LJ}(p_0,p_0;E),
\end{equation}
where $\mu(p_0)$ is the relativistic reduced mass, given by \cite{vorabbi2016}
\begin{equation}\label{eq36}
\mu(p_0)=\frac{\sqrt{(p_0^2+m^2)(p_0^2+M_t^2)}}{\sqrt{p_0^2+m^2}+\sqrt{p_0^2+M_t^2}}.
\end{equation}
Here $m$ and $M_t$ are the incident nucleon and struck nucleus masses, respectively.

The unpolarized differential cross section, the probability of scattering into a small solid angle $d\Omega$ averaged over initial spin states and summed over final spin states. The standard result for the spin-1/2 unpolarized elastic scattering differential cross section is
\begin{equation}\label{eq37}
\frac{d\sigma}{d\Omega}=\vert B_c(p_0,\Theta)\vert^2+\vert C(p_0,\Theta)\vert^2.
\end{equation}
The analyzing power $A_y$ is calculated using the formula
\begin{equation}\label{eqAy}
A_y=\frac{2\operatorname{Re}\left[B_c^\ast(p_0,\Theta)C(p_0,\Theta)\right]}{\vert B_c(p_0,\Theta)\vert^2+\vert C(p_0,\Theta)\vert^2}.
\end{equation}
Whereas the spin rotation $Q$ is calculated using the formula
\begin{equation}\label{eqQ}
Q=\frac{2\operatorname{Im}\left[B_c(p_0,\Theta)C^\ast(p_0,\Theta)\right]}{\vert B_c(p_0,\Theta)\vert^2+\vert C(p_0,\Theta)\vert^2}.
\end{equation}

\subsection{Coulomb Correction for $pA$ scattering}

Various techniques exist to account for the Coulomb interaction in $pp$ or $pA$ scattering, including incorporating the Coulomb potential into the momentum-space Watson multiple-scattering expansion \cite{chinn1991,elster1993,kume1999}.

In this work, the nuclear optical $NN$ scattering amplitudes do not include the point-Coulomb amplitude. Instead, the amplitudes are refined in Eqs.(\ref{eq33}) and (\ref{eq34}) by incorporating the Coulomb-distortion
phase $e^{2i\sigma_L}$ into each partial wave, where the Coulomb phases $\sigma_L$ are defined as
\begin{equation}\label{eqColphase}
\sigma_L = \arg\Gamma(1+L+i\eta).
\end{equation}
 Here, $\eta = \alpha/\beta_{cm}$. The the point-Coulomb (Rutherford) amplitude is implemented in the CM frame for proton-nucleus ($pA$) scattering off a spin-0 nucleus, given by
\begin{equation}\label{eqCoulomb}
f_c(\Theta) = -\frac{\eta}{2k_{pA}\sin^2(\Theta/2)} \exp\big[-i\eta\ln\sin^2(\Theta/2) + 2i\sigma_0\big],
\end{equation}
where $k_{pA}$ represents the center-of-mass (c.m.) $pA$ wave number, and $\Theta$ is the $pA$ c.m. scattering angle. The $pA$ scattering angle $\Theta$ is related to the $NN$ scattering angle $\theta$ via
the invariant momentum transfer $q = \vert\mathbf{p}^\prime-\mathbf{p}\vert$.

The differential cross-section given in eq.(\ref{eq37}) is modified to be
\begin{equation}\label{eq37Coulomb}
\frac{d\sigma}{d\Omega}=\vert B^d_c(\Theta)+f_C(\Theta)\vert^2+\vert C^d(\Theta)\vert^2,
\end{equation}
where $B^d_c$ and $C^d$ are the distorted central and spin-orbit amplitudes, respectively. The analyzing power eq.(\ref{eqAy}) and the spin rotation  eq.(\ref{eqQ}) are modified to
\begin{equation}\label{eqAyC}
A_y=\frac{2\operatorname{Re}\left\{\left[B_c^d(\Theta)+f_C(\Theta)\right]^\ast C(\Theta)\right\}}{\vert B_c^d(\Theta)\vert^2+\vert C^d(\Theta)\vert^2},
\end{equation}
and
\begin{equation}\label{eqQC}
Q=\frac{2\operatorname{Im}\left\{\left[B_c^d(\Theta)+f_C(\Theta)\right]C^\ast(\Theta)\right\}}{\vert B_c^d(\Theta)\vert^2+\vert C^d(\Theta)\vert^2}.
\end{equation}

\subsection{The $NN$ amplitudes}
The $NN$ elastic scattering amplitude for the scattering from incoming momentum $\mathbf{p}$ to outgoing momentum $\mathbf{p}'$ is denoted by $M(\mathbf{p}',\mathbf{p},\mathcal{E})$ and deﬁned as follows \cite{vorabbi2016,crespo1992}
\begin{equation}\label{eq38}
\mathcal{M}(\mathbf{p}',\mathbf{p},\mathcal{E})=-4\pi^2\mu_{\mathrm{NN}}\langle\mathbf{p}'\vert t(\mathbf{p}',\mathbf{p},\mathcal{E})\vert\mathbf{p}\rangle,
\end{equation}
where $\mu_\mathrm{NN}$ is the $NN$ reduced mass, $t$ is the $NN$ $t$-matrix given in defined in eq.(\ref{eq20}) and (\ref{eq23}). Let us define the normal $\hat{\mathbf{n}}$, longitudinal $\hat{\mathbf{P}}$, and transverse $\hat{\mathbf{q}}$ momentum components with respect to the scattering plane as follows \cite{vorabbi2016,baker2022,hoshizaki1968}:
\begin{equation}\label{eq39}
\hat{\mathbf{n}}=\frac{\mathbf{p}\times\mathbf{p}'}{\vert\mathbf{p}\times\mathbf{p}'\vert};\quad
\hat{\mathbf{P}}=\frac{\mathbf{p}'+\mathbf{p}}{\vert\mathbf{p}'+\mathbf{p}\vert};\quad
\hat{\mathbf{q}}=\frac{\mathbf{p}'-\mathbf{p}}{\vert\mathbf{p}'-\mathbf{p}\vert}.
\end{equation}
Note that $\hat{\mathbf{q}}=\hat{\mathbf{n}}\times\hat{\mathbf{P}}$ \cite{hoshizaki1968}. The most general form of this amplitude, consistent with invariance under rotation, time reversal, and parity is \cite{wolfenstein1954}
\begin{eqnarray}
\mathcal{M}&=&B\,\mathbf{S}+C(\bm{\sigma}+\bm{\sigma}_t)\bm{\cdot}\hat{\mathbf{n}}\notag\\
&+&\frac12G\left[(\bm{\sigma\cdot}\hat{\mathbf{q}})(\bm{\sigma}_t\bm{\cdot}\hat{\mathbf{q}})+(\bm{\sigma\cdot}\hat{\mathbf{P}})(\bm{\sigma}_t\bm{\cdot}\hat{\mathbf{P}})\right]\mathbf{T}\notag\\
&+&\frac12H\left[(\bm{\sigma\cdot}\hat{\mathbf{q}})(\bm{\sigma}_t\bm{\cdot}\hat{\mathbf{q}})-(\bm{\sigma\cdot}\hat{\mathbf{P}})(\bm{\sigma}_t\bm{\cdot}\hat{\mathbf{P}})\right]\mathbf{T}\notag\\
&+&N(\bm{\sigma\cdot}\hat{\mathbf{n}})(\bm{\sigma}_t\bm{\cdot}\hat{\mathbf{n}})\mathbf{T},\label{eq40}
\end{eqnarray}
where $\mathbf{S}$ and $\mathbf{T}$ are singlet and triplet operators, respectively. $\bm{\sigma}$ and $\bm{\sigma}_t$ are the spin operator for the incident and the target nucleon. From parity considerations it follows that $B$ and $H$ are even functions of $\cos\theta$, $G$ and $N$ are odd functions, and $C$ is an even function times $\sin\theta$. Comparing Eq.(\ref{eq40}) with Eq.(\ref{eq32}) one can find the operators $B_c$ and $C_s$ to obtain
\begin{eqnarray}\label{eq41}
\frac{d\sigma}{d\Omega}&=&\frac14 \mathrm{Tr}(\mathcal{M}\mathcal{M}^\dagger)=\frac14\vert B\vert^2+2\vert C\vert^2+\notag\\
&&\frac14\vert G-N\vert^2+\frac12\vert N\vert^2+\frac12\vert H\vert^2.
\end{eqnarray}
The amplitudes $B$, $C$, $G$, $H$, and $N$ are called the Wolfenstein amplitudes \cite{wolfenstein1954,McClung2025,burrows2020}, and they can be expressed as complex functions of $\mathbf{p}$, $\mathbf{p}'$, and $\mathcal{E}$.

The $NN$ amplitudes are usually expressed in terms of the decomposition of the scattering amplitude into components describing spin singlet ($S = 0$) and spin triplet ($S = 1$)
scattering, $\mathcal{M}^S_{m'_S,m_S}$, where $m_S$ and $m'_S$ refer to the incident- and
ﬁnal-spin projections in the triplet state \cite{vorabbi2016,crespo1992}. In the representation in which these projections are referred to an axis of quantization along the incident beam direction ($\hat{\mathbf{p}}$) one has \cite{vorabbi2016,crespo1992}
\begin{equation}\label{eqBWolf}
B = \frac{1}{4} \left( 2\mathcal{M}^1_{11} + \mathcal{M}^1_{00} + \mathcal{M}^0_{00} \right),
\end{equation}
\begin{equation}\label{eqCWolf}
C = \frac{1}{2\sqrt{i}} \left( \mathcal{M}^1_{10} - \mathcal{M}^1_{01} \right),
\end{equation}
\begin{equation}\label{eqGHWolf}
G+H = \frac{1}{4} \left( \mathcal{M}^1_{00} - 2\mathcal{M}^1_{1,-1} - \mathcal{M}^0_{00} \right).
\end{equation}
The rest of the parameters can be deduced from refs.\cite{vorabbi2016,crespo1992}. The amplitudes $\mathcal{M}^S_{m'_S,m_S}$ are obtained in terms of the partial-wave components of the $NN$ amplitude, $\mathcal{M}^{JS}_{L'L}(p',p,\mathcal{E}$, defined as \cite{vorabbi2016,crespo1992,paez1984}
\begin{equation}\label{eqPWEM}
\begin{split}
\mathcal{M}(\mathbf{p}', \mathbf{p}; \mathcal{E})=\frac2\pi\sum_{JLL'SM}\,&i^{L-L'}\mathscr{Y}^{L'S}_{JM}(\mathbf{\hat{p}}')\times\\
& \mathcal{M}^{JS}_{L'L}(p', p; \mathcal{E}){\mathscr{Y}^{LS}_{JM}}^\dagger(\mathbf{\hat{p}}),
\end{split}
\end{equation}
where ${\mathscr{Y}^{LS}_{JM}}(\mathbf{\hat{p}})$ is similar to eq.(\ref{eq26}) but for spin-$S$, defined by coupling orbital and spin degrees of freedom
\begin{equation}\label{eq26S}
\mathscr{Y}^{LS}_{JM}(\cos\theta)=\sum_{m_L,m_s}\langle Lm_L\, S m_S|JM\rangle Y_L^{m_L}(\cos\theta)\chi_{S}^{m_s}.
\end{equation}

The amplitude $\mathcal{M}^S_{m'_S,m_S}$ is expressed explicitly as
\begin{equation}\label{eqMsms}
\begin{split}
\mathcal{M}^S_{m'_S, m_S} = \frac{2}{\pi}&\sum_{J,M}\sum_{\substack{ L,L'\\ m_{L},m'_{L}}} i^{L-L'}\times \\
&\langle L' m'_L, S m'_S \mid J M \rangle\langle L m_L, S m_S \mid J M \rangle\times\\ 
&Y_{L'}^{m'_L}(\hat{p}') \, Y_{L}^{m_L \ast}(\hat{p}) \, \mathcal{M}_{L'L}^{JS}(p', p; \mathcal{E}).
\end{split}
\end{equation}
Expressing the $\mathcal{M}^S_{m'_S,m_S}$ amplitudes in terms of the partial-wave amplitudes $\mathcal{M}_{L'L}^{JS}(p', p; \mathcal{E})$ for a
quantization axis along the incident-beam direction can be found in Refs.\cite{crespo1992, paez1984}. The $B$ and $C$ amplitudes are given by \cite{vorabbi2016}
\begin{equation}\label{eqWB}
\begin{split}
B_{pN}=&\frac{1}{4\pi ^{2}}\sum_{L=0}^{\infty}P_{L}(\cos\theta)\times \\
&\left[\begin{array}{c}
(2L+1)\mathcal{M}_{LL}^{L,S=0}+(2L+1)\mathcal{M}_{LL}^{L,S=1}+ \\\\ 
(2L+3)\mathcal{M}_{LL}^{L+1,S=1}+(2L-1)\mathcal{M}_{LL}^{L-1,S=1}%
\end{array}\right],
\end{split}
\end{equation}
and
\begin{equation}\label{eqWC}
\begin{split}
C_{pN}=&\frac{i}{4\pi ^{2}}\sum_{L=1}^{\infty}P_{L}^{1}(\cos\theta)\times \\
&\left[\begin{array}{c}
\left( \frac{2L+3}{L+1}\right) \mathcal{M}_{LL}^{L+1,S=1}- \\ \\ 
\left( \frac{2L+1}{L(L+1)}\right) \mathcal{M}_{LL}^{L,S=1}-\left( \frac{2L-1%
}{L}\right) \mathcal{M}_{LL}^{L-1,S=1}%
\end{array}\right].
\end{split}
\end{equation}
The Coulomb distortion phase factors $\exp(2i\sigma_L)$, where $\sigma_L$ is defined in Eq. (\ref{eqColphase}), have been incorporated into the summations for the $B_{pp}$ and $C_{pp}$ amplitudes
(eqs. \ref{eqWB} and \ref{eqWC}). While previous studies by refs \cite{vorabbi2016,burrows2020,McClung2025} exhibited deviations from experimental values, the inclusion of these factors significantly improves the realism
of the amplitudes, particularly for $\operatorname{Re}C_{pp}$, leading to better agreement with experimental data.

An explicit expression for the $pp$ and $pn$ central and spin-orbit parts of the $NN$ $t$-matrix given in eq.(\ref{eq23}) in terms of the partial-wave components $t^{ST}_{JLL}(p',p;\mathcal{E})$
are given for $pp$ central part \cite{vorabbi2016}
\begin{equation}\label{eqtcpp}
\begin{split}
t_{pp}^{c}=&\frac{1}{4\pi ^{2}}\sum_{L=0}^{\infty }P_{L}(\cos\theta)\times\\
&\left[\begin{array}{c}
(2L+1)t_{L,LL}^{S=0,T=1}+(2L+1)t_{L,LL}^{S=1,T=1}+ \\\\ 
(2L-1)t_{L-1,LL}^{S=1,T=1}+(2L+3)t_{L+1,LL}^{S=1,T=1}%
\end{array}%
\right],
\end{split}
\end{equation}
and for $pp$ spin-orbit part
\begin{equation}\label{eqtlspp}
\begin{split}
t_{pp}^{ls}=&-\frac{1}{2\pi ^{2}(p^{\prime }p)}\sum_{L=1}^{\infty}
\frac{dP_{L}(\cos \theta )}{d(\cos\theta)}\times\\
&\left[\begin{array}{c}
-\frac{2L-1}{L}t_{L-1,LL}^{S=1,T=1}-\frac{2L+1}{L(L+1)}t_{L,LL}^{S=1,T=1}\\ \\ 
+\frac{2L+3}{L+1}t_{L+1,LL}^{S=1,T=1}%
\end{array}\right].
\end{split}
\end{equation}
For $pn$ central part \cite{vorabbi2016}
\begin{equation}\label{eqtcpn}
\begin{split}
t_{pn}^{c}=&\frac{1}{8\pi ^{2}}\sum_{L=0}^{\infty }
P_{L}(\cos\theta)\times \\
&\left[\begin{array}{c}
(2L+1)t_{L,LL}^{S=0,T=0}+(2L+1)t_{L,LL}^{S=1,T=0}+ \\ \\ 
(2L-1)t_{L-1,LL}^{S=1,T=0}+(2L+3)t_{L+1,LL}^{S=1,T=0}+ \\ \\ 
(2L+1)t_{L,LL}^{S=0,T=1}+(2L+1)t_{L,LL}^{S=1,T=1}+ \\ \\
(2L-1)t_{L-1,LL}^{S=1,T=1}+(2L+3)t_{L+1,LL}^{S=1,T=1}%
\end{array}\right],
\end{split}
\end{equation}
and for $pn$ spin-orbit part
\begin{equation}\label{eqtlspn}
\begin{split}
t_{pn}^{ls}=&-\frac{1}{4\pi ^{2}(p^{\prime }p)}\sum_{L=1}^{\infty }
\frac{dP_{L}(\cos \theta )}{d(\cos \theta )}\times\\
&\left[\begin{array}{c}
-\frac{2L-1}{L}t_{L-1,LL}^{S=1,T=0}-\frac{2L+1}{L(L+1)}t_{L,LL}^{S=1,T=0}+\\\\ 
\frac{2L+3}{L+1}t_{L+1,LL}^{S=1,T=0}-\frac{2L-1}{L}t_{L-1,LL}^{S=1,T=1}- \\ \\
\frac{2L+1}{L(L+1)}t_{L,LL}^{S=1,T=1}+\frac{2L+3}{L+1}t_{L+1,LL}^{S=1,T=1}%
\end{array}\right].
\end{split}
\end{equation}
The partial-wave components $t^{ST}_{JLL}(p',p;\mathcal{E})$ are computed in the $NN$ center-of-mass frame for each partial wave up to $J = 8$, using the EGM $NN$ potential \cite{epelbaum2005}.
Computation procedures follow the exact recipes of ref. \cite{vorabbi2016}.

\subsection{The chiral potential}
In $\chi^\mathrm{EFT}$, the different contributions to nuclear forces are arranged according to their importance by employing a power-counting scheme \cite{epelbaum2005, machleidt2011}. The contact $NN$ potential is then given as a series of terms
\begin{equation}\label{eq42}
V_\chi=V^{(0)}+V^{(2)}+V^{(4)}+\cdots,
\end{equation}
The superscript labels the order in the chiral expansion, which is organized in powers of the ratio $Q/\Lambda_\chi$. Here $Q$ is the soft scale — representing either typical nucleon momenta or the pion mass $m_\pi$ — while $\Lambda_\chi\approx m_\rho$ is the hard scale (roughly 0.5–1 GeV), the momentum at which the chiral expansion is expected to break down
\cite{machleidt2011}.

All contributions through N$^3$LO are included. These contributions fall into two categories:
\begin{itemize}
\item Long- and intermediate-range contributions arise from the exchange of one or more pions. These are fully determined by the chiral symmetry of QCD together with low-energy pion-nucleon scattering data -- no free parameters remain once the $\pi N$ system is characterized.
\item Contact (short-range) contributions encode physics at distances too short for the pion-exchange picture to apply. They are parametrized by low-energy constants (LECs) --- coefficients that must be fit to experimental nucleon-nucleon scattering data, since chiral symmetry alone does not fix them \cite{epelbaum2005, machleidt2011}.
\end{itemize}

Calculations are carried out using the nucleon–nucleon chiral interaction at fourth order (N$^3$LO), developed by E. Epelbaum, W. Gl\"ockle, and Ulf-G. Mei\ss ner \cite{epelbaum2005}. In this formulation, commonly referred to as the EGM potential, the two-pion exchange (2PE) contributions are treated using the spectral function regularization (SFR) scheme. The spectral function regulator is implemented by replacing the upper limit of the pion-loop integral with a finite cutoff $\tilde{\Lambda}$, rather than taking it to infinity. The purpose of this regularization is to suppress the short-range components of the 2PE interaction, which otherwise generate an unrealistically strong attractive force at high momenta.

To further control the high-momentum behavior and ensure the convergence of the scattering equation, the $NN$ potential entering the LS equation is multiplied by a Gaussian regulator function $f_\Lambda(p)$ given by \cite{epelbaum2005}
\begin{equation}\label{eq43}
f_\Lambda(p)= \exp\left[ -\left(\frac{p}{\Lambda}\right)^{2n}\right],
\end{equation}
where $n$ is integer chosen to be large enough to effectively removes momentum components above a cutoff scale $\Lambda$, thereby restricting the interaction to the low-energy region where chiral effective field theory is applicable and reliable. typically, $n=2$ or $3$, yielding a "super-Gaussian" to keep low-energy components unperturbed. The regularized potential is evaluated as
\begin{equation}
V_\mathrm{reg}(\mathbf{p}', \mathbf{p}) = f_\Lambda(p') \, V(\mathbf{p}', \mathbf{p}) \, f_\Lambda(p).
\end{equation}
make use in Eq.(\ref{eq3}), yields
\begin{equation}
T(\mathbf{p}', \mathbf{p}; E) = V_\mathrm{reg}(\mathbf{p}', \mathbf{p}) + \int \frac{d^3k}{(2\pi)^3}\frac{V_\mathrm{reg}(\mathbf{p}', \mathbf{k})T(\mathbf{k}, \mathbf{p}; E)}{E - \frac{k^2}{2\mu} + i\epsilon}.
\label{eq3reg}
\end{equation}
In this work $n = 3$ \cite{epelbaum2005}.  As the intermediate momentum   removing UV divergences and ensuring numerical convergence. The low-energy constants (LECs) depend implicitly on the regularization scheme and the cutoff. The calculated observables should be independent of the employed cutoff to ensure the renormalization group invariance (RGI).

\section{Results and discussions}\label{sec:results}
The Wolfenstein amplitudes are evaluated using the $\chi^\mathrm{EFT}$ $t$-matrix and optical potential, employing both the Gaussian regulator and EGM's SFR scheme with $(\Lambda,\tilde{\Lambda}) = (450, 500)$, $(550, 600)$, and $(600, 600)$ MeV, for $pp$ and $pn$ scattering at 100 and 200 MeV \cite{epelbaum2005}. The observables--the differential cross section, analyzing power, and spin rotation-- are computed for \ce{p + ^{16}O} and \ce{p + ^{40}Ca} elastic scattering at 100 and 200 MeV.

\subsection{The Wolfenstein amplitudes}
\begin{figure*}[htbp]
\centering
\includegraphics[width=0.99 \textwidth]{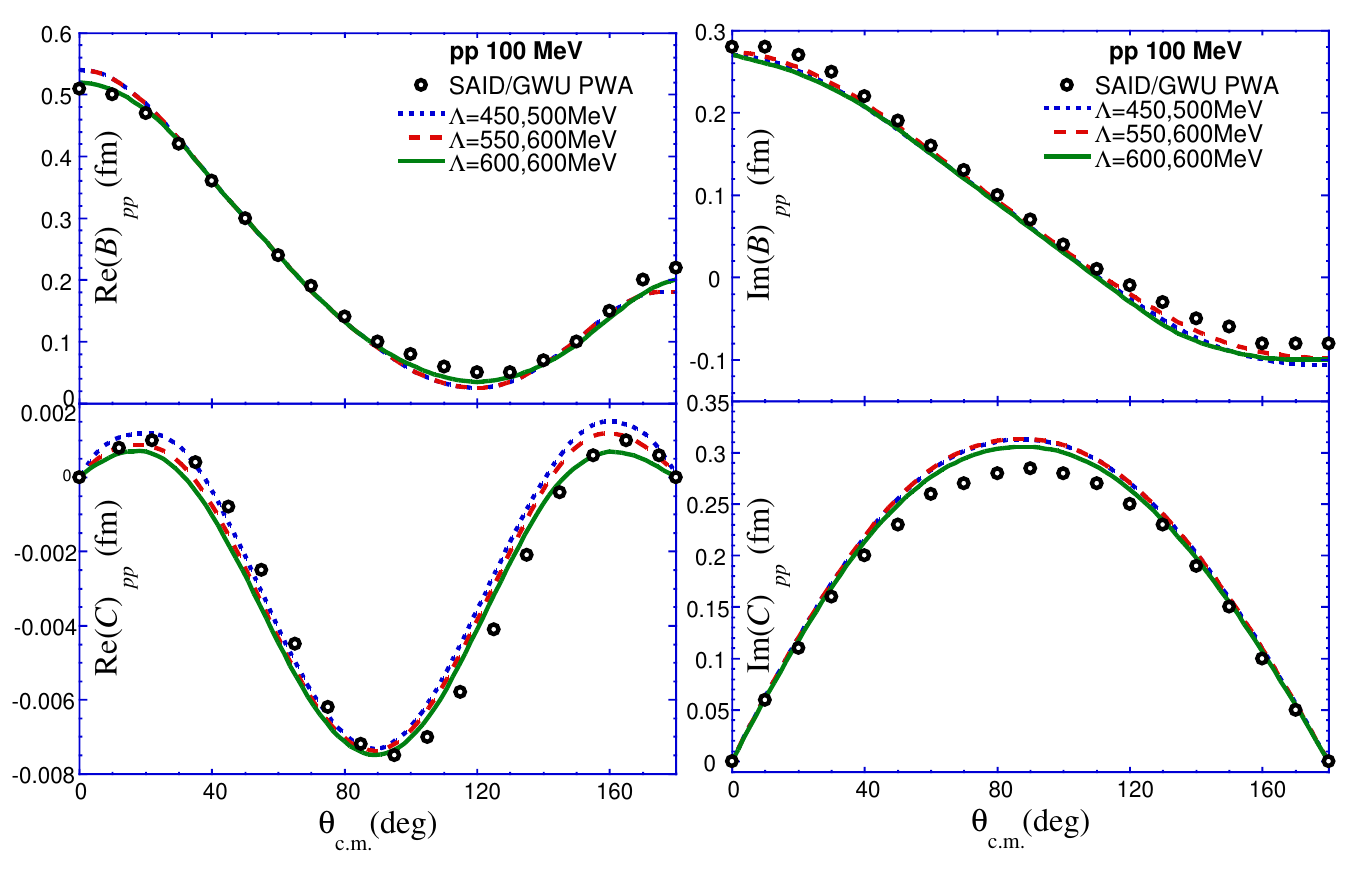}
\caption{Real (left) and imaginary (right) parts of the central Wolfenstein amplitude $B_{pp}$ (top) and the spin-orbit amplitude $C_{pp}$ (bottom) for $pp$ scattering at 100 MeV, as functions of the center-of-mass angle $\theta_{cm}$, with all amplitudes in fm. Curves show the $\chi^\mathrm{EFT}$ optical-potential predictions for the cutoff values shown in legends. Black circles: amplitudes reconstructed from the SAID/GWU partial-wave analysis \cite{said_sm16}.}
\label{figWApp100}
\end{figure*}

Figure \ref{figWApp100} compares experimental data (black circles) with the theoretical predictions for the central (spin-independent) Wolfenstein amplitude $B_{pp}$ and the spin-orbit amplitude $C_{pp}$ in $pp$ scattering at 100 MeV, plotted as functions of the center-of-mass scattering angle $\theta_{cm}$. The left column shows the real parts ($\operatorname{Re}B_{pp}$ top, $\operatorname{Re}C_{pp}$ bottom) and the right column the imaginary parts ($\operatorname{Im}B_{pp}$ top, $\operatorname{Im}C_{pp}$ bottom), with all amplitudes given in fm. The calculations are performed for the various cutoff values specified earlier.
The reference amplitudes were reconstructed from the SAID partial-wave analysis \cite{said_sm16}, using the phase shifts available from the SAID database \cite{said_web}.

The results presented in Figure \ref{figWApp100} highlight the convergence properties of the calculation at 100 MeV. At N$^3$LO, the calculation reproduces the real and imaginary parts of amplitudes $B_{pp}$ and $C_{pp}$ within a narrow regulator-dependence band, which is consistent with the expected $\mathcal{O}(Q^5)$ truncation uncertainty. This narrowness serves as evidence that the $\chi^\mathrm{EFT}$ is well-converged at this energy, as a wider cut-off band at this order would typically signal incomplete convergence or spurious artifacts.

The spin-orbit amplitude $C_{pp}$ shows excellent agreement with experimental data in terms of sign, magnitude, and peak position. Note that the residual regulator dependence is minimized by the choice of the softer regulator combination, $(\Lambda,\tilde{\Lambda}) = (450, 500)$ MeV. While the harder regulators, particularly $(\Lambda,\tilde{\Lambda}) =(600, 600)$ MeV, remain physically consistent, the softer choice provides the cleanest convergence profile, avoiding potential finite-cut-off artifacts that can arise at higher orders.

\begin{figure*}[htbp]
\centering
\includegraphics[width=0.99 \textwidth]{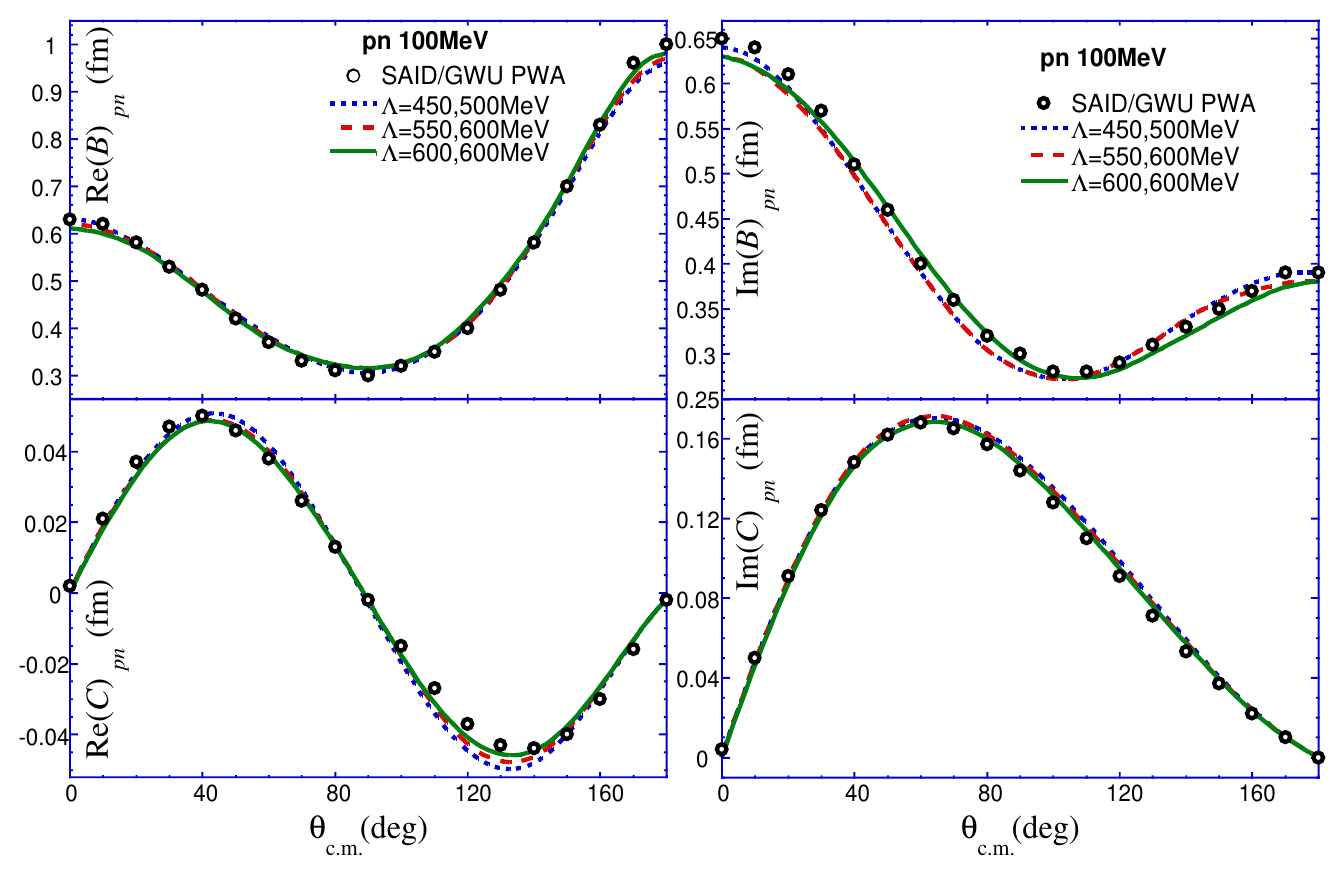}
\caption{Real (left) and imaginary (right) parts of the central Wolfenstein amplitude $B_{pn}$ (top) and the spin-orbit amplitude $C_{pn}$ (bottom) for $pn$ scattering at 100 MeV, as functions of the center-of-mass angle $\theta_{cm}$, with all amplitudes in fm. Curves show the $\chi^\mathrm{EFT}$ optical-potential predictions for the cutoff values shown in legends. Black circles: amplitudes reconstructed from the SAID/GWU partial-wave analysis \cite{said_sm16}.}
\label{figWApn100}
\end{figure*}

Figure \ref{figWApn100} is similar to figure \ref{figWApp100}, but for the pn scattering amplitudes at 100 MeV. The figure demonstrates the high predictive power and stability of the framework at N$^3$LO. The agreement between the theoretical calculations and experimental data is excellent, with the calculated real and imaginary parts of amplitudes $B_{pn}$ and $C_{pn}$ accurately tracking the data points across the entire angular range. Notably, the model successfully reproduces the kinematic zeros of the spin-orbit amplitude $C_{pn}$ at both $0^\circ$ and $180^\circ$, confirming that the framework correctly respects the necessary symmetry constraints at the angular extremes.

Furthermore, the results exhibit a remarkably constrained regulator dependence, indicating that the $\chi^\mathrm{EFT}$ is well-converged at this energy. The grouping of the $(\Lambda,\tilde{\Lambda })$  $(450, 500)$, $(550, 600)$, and $(600, 600)$ MeV bands is tightly clustered, which is consistent with the expected $\mathcal{O}(Q^5)$ truncation uncertainty. The residual regulator dependence is minimized by the choice of the softer (450,500) MeV regulator, providing a clean convergence profile that underscores the robustness of the current N$^3$LO expansion in the pn channel.

\begin{figure*}[htbp]
\centering
\includegraphics[width=0.99 \textwidth]{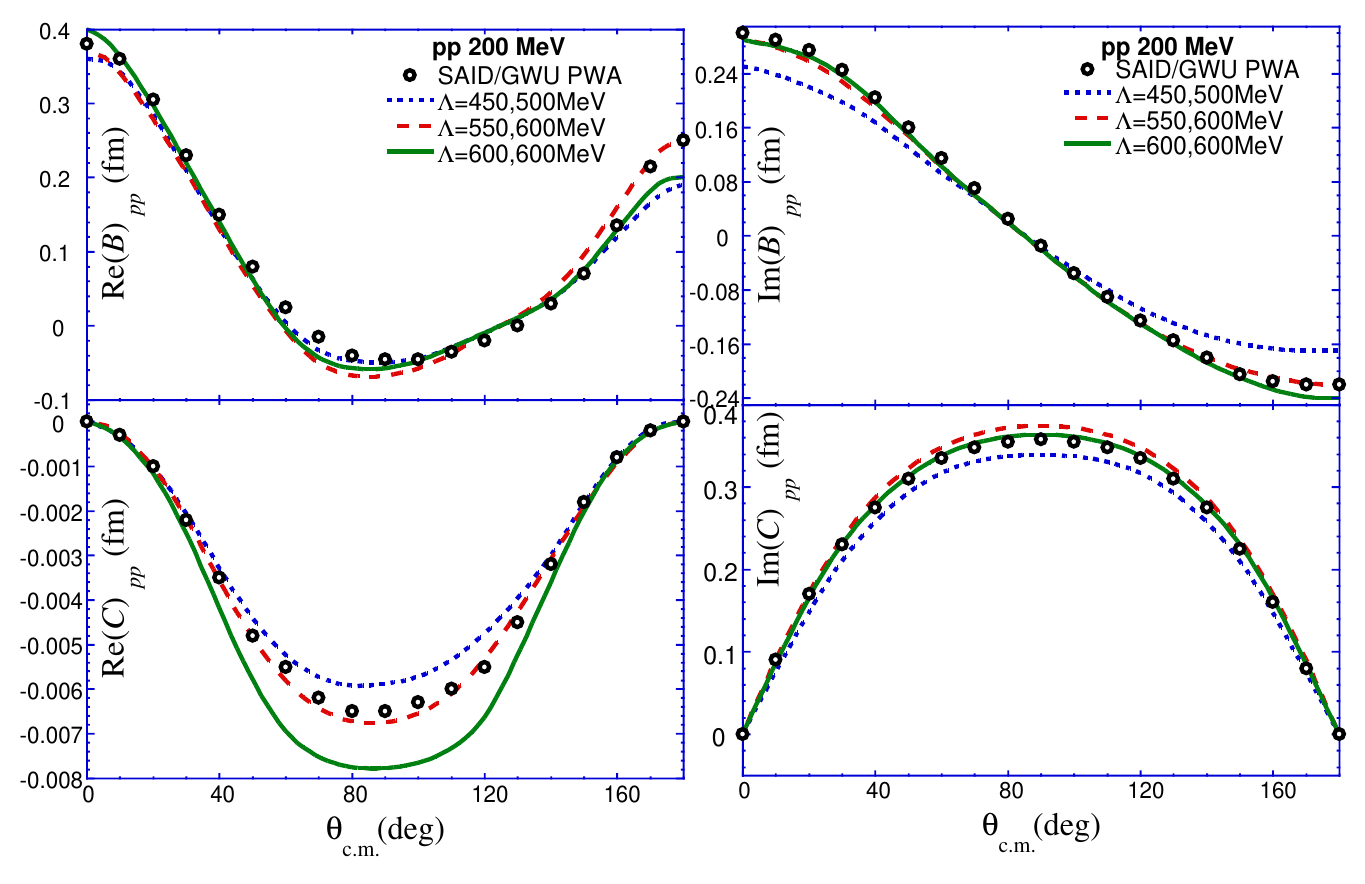}
\caption{Real (left) and imaginary (right) parts of the central Wolfenstein amplitude $B_{pp}$ (top) and the spin-orbit amplitude $C_{pp}$ (bottom) for $pp$ scattering at 200 MeV, as functions of the center-of-mass angle $\theta_{cm}$, with all amplitudes in fm. Curves show the $\chi^\mathrm{EFT}$ optical-potential predictions for the cutoff values shown in legends. Black circles: amplitudes reconstructed from the SAID/GWU partial-wave analysis \cite{said_sm16}.}
\label{figWApp200}
\end{figure*}

\begin{figure*}[htbp]
\centering
\includegraphics[width=0.99 \textwidth]{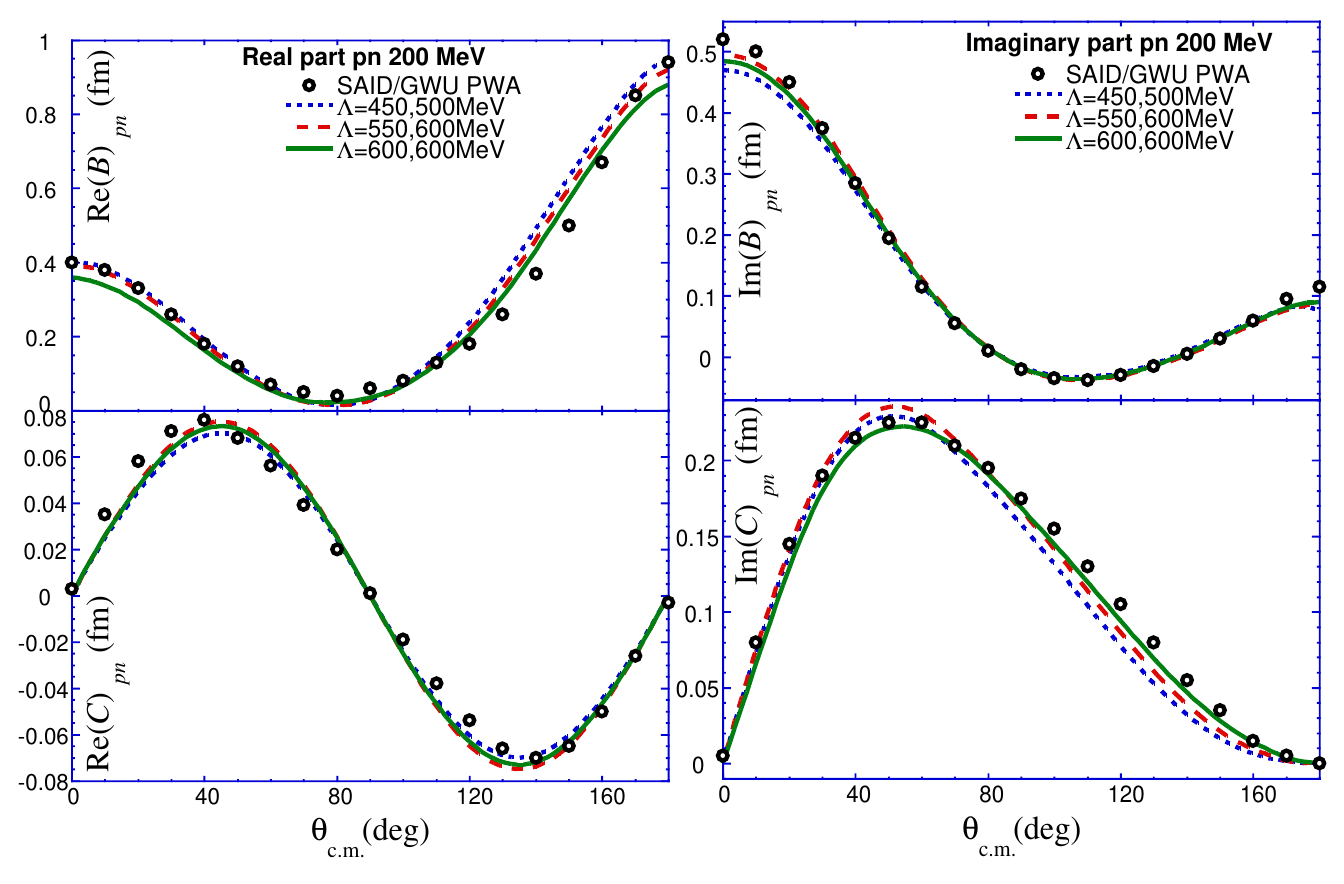}
\caption{Real (left) and imaginary (right) parts of the central Wolfenstein amplitude $B_{pn}$ (top) and the spin-orbit amplitude $C_{pn}$ (bottom) for pn scattering at 200 MeV, as functions of the center-of-mass angle $\theta_{cm}$, with all amplitudes in fm. Curves show the $\chi^\mathrm{EFT}$ optical-potential predictions for the cutoff values shown in legends. Black circles: amplitudes reconstructed from the SAID/GWU partial-wave analysis \cite{said_sm16}.}
\label{figWApn200}
\end{figure*}

Figures \ref{figWApp200} and \ref{figWApn200} show the same results as \ref{figWApp100} and \ref{figWApn100}, respectively, but for a proton lab energy of 200 MeV. At 200 MeV the agreement between theoretical predictions and experiment deteriorates relative to the 100 MeV case for both reactions, with the degradation most pronounced for pp scattering. The cutoff bands broaden and the curves depart more visibly from the data, particularly in the spin-orbit amplitude $C_{pp}$ and in the imaginary parts.


The loss of agreement is expected and has a clear EFT origin, reinforced by reaction-specific dynamics. The chiral expansion is controlled by $Q \approx \frac{p}{\Lambda_\chi}$ and the relative momentum at 200 MeV is larger than at 100 MeV (roughly a factor $\sqrt{2}$ in $p$ for non-relativistic kinematics). Since the leading omitted term at N$^3$LO scales as $Q^5$, the truncation error grows steeply -- a modest increase in $Q$ produces a much larger increase in $Q^5$. So even at fixed chiral order, the theory is intrinsically less accurate at 200 MeV, and the widening cutoff band you see is the direct visual signature of this enlarged truncation uncertainty.

At 200 MeV the probed momenta sit closer to $\Lambda_\chi \approx 600$ MeV, so regulator artifacts grow and the residual cutoff dependence -- which should be mild at low energy -- becomes more visible. This is why the three cutoff combinations now spread apart instead of clustering, and why the harder regulators, especially (600,600), deviate more.

Another reason for such deterioration is that inelasticity is starting to kick in. The pion-production threshold sits near $K_{\text{lab}} \approx 280$ MeV, so at 200 MeV the system is approaching it. Absorptive/inelastic effects become more important, and these are precisely the channels that the optical potential treats most phenomenologically. This is why the imaginary parts degrade more than the real parts -- $\operatorname{Im}B$ and $\operatorname{Im}C$ carry the absorption physics that is least well constrained at this energy.

The $pp$ suffers more than $pn$. A few reasons combine: 
\begin{itemize}
\item The $pp$ amplitudes are smaller in magnitude at 200 MeV ($\operatorname{Re}B_{pp}$ runs only ~0.0–0.4 and $\operatorname{Re}C_{pp}$ is order $10^{-3}$ fm, so the same absolute theory error translates into a larger {\em relative} discrepancy than in $pn$, where the amplitudes are several times larger ($\operatorname{Re}B_{pp}$ up to $\approx 1$).
\item The $pp$ system is restricted to isospin $T=1$ only, so it lacks the $T=0$ partial waves (notably the strong coupled $^3S_1$--$^3D_1$ channel) that dominate and stabilize the $pn$ amplitudes. The observables are therefore more sensitive to the finer, less-constrained pieces of the interaction.
\item Coulomb–nuclear interference and the electromagnetic treatment add an extra layer that is harder to control as energy rises, affecting pp but not $pn$.
\end{itemize}

\subsection{The proton-nucleus scattering}
\begin{figure}[htbp]
\centering
\includegraphics[width=0.99 \linewidth]{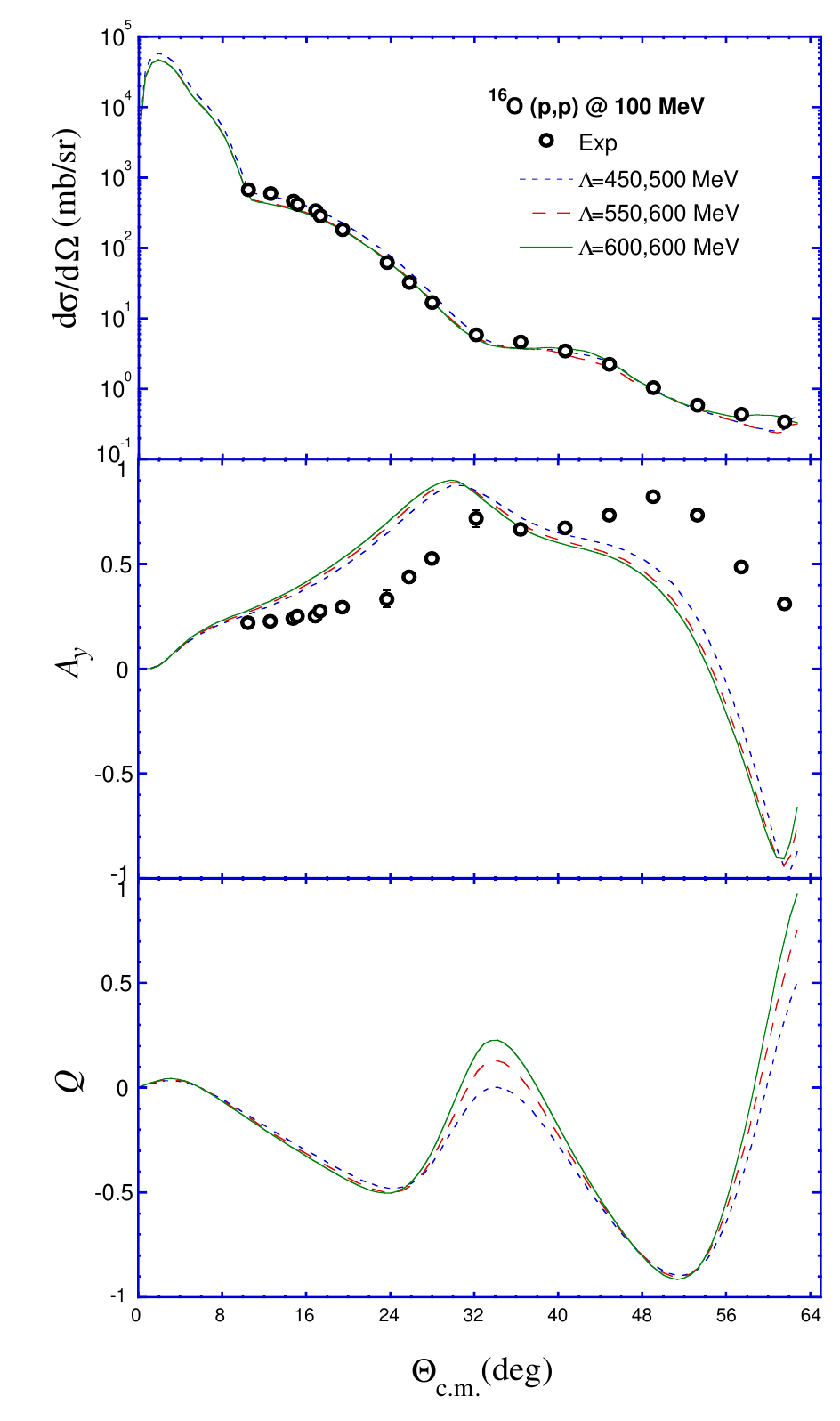}
\caption{Elastic scattering observables for 100 MeV protons on \ce{^{16}O}. Comparison of differential cross section ($d\sigma/d\Omega$), analyzing power ($A_y$), and spin rotation parameter ($Q$)
calculated using a single-folding optical potential based on N$^3$LO $\chi^\mathrm{EFT}$. Calculations utilize regulator cutoffs $(\Lambda,\tilde{\Lambda })$ of $(450, 500)$, $(550, 600)$, and $(600, 600)$ MeV.
Black circles are experimental data taken from the data tables of \cite{KellyNA} which compile the results of Ref.\cite{seifert1990}; cross-section uncertainties are smaller than the symbols.}
\label{O16_100}
\end{figure}

In this study, the elastic scattering of protons by \ce{^{16}O} and \ce{^{40}Ca} is examined at laboratory energies ranging from 100 to 200 MeV. Figure \ref{O16_100} shows the elastic scattering of protons from $^{16}\text{O}$ at 100 MeV. The calculations were performed across a range of regulator cutoff values $\Lambda$, spanning from soft (450, 500) MeV to hard (600, 600 MeV) configurations, to assess the sensitivity of the results to short-range physics and the theoretical uncertainty of the $\chi^\mathrm{EFT}$ expansion. The results are presented for three primary observables: the differential cross section ($d\sigma/d\Omega$), the analyzing power ($A_y$), and the spin rotation parameter ($Q$).

The differential cross section serves as the most robust diagnostic of the calculation, with the forward diffraction pattern ($\Theta_{\text{cm}} \lesssim 40^\circ$) the cross section is reproduced well over the
measured angular range across all cutoff values. This region, dominated by the nuclear surface and gross geometry, exhibits minimal dependence on the regulator, indicating that the long-range physics is
well-captured. As expected, the theoretical uncertainty—represented by the spread of the three curves—increases at backward angles, where higher momentum transfer probes short-distance physics sensitive
to the $\chi^\mathrm{EFT}$ regulator. The observed hierarchy, where softer cutoffs result in smoother curves and harder cutoffs exhibit more pronounced oscillatory structure, confirms the expected influence
of the regulator on high-momentum strength.

The analyzing power $A_y$ and the spin rotation parameter $Q$ provide more stringent tests of the spin-dependent components of the interaction. The analyzing power $A_y$ is overestimated below about $30^\circ$,
and above about $40^\circ$ the data remain at 0.7-0.8 whereas the calculation decreases. For $A_y$, the observed discrepancy between the calculations and
experimental data--particularly the underprediction of the magnitude at oscillation maxima--is a well-documented phenomenon in microscopic $\chi^\mathrm{EFT}$-based folding models, despite the inclusion of the
three-nucleon force (3NF) contributions and medium-dependent modifications to the spin-orbit interaction. These differences are mainly attributed to the first-order truncation of the optical potential,
which does not include medium effects such as Pauli blocking and density dependence of the effective interaction. These effects are known to be significant for proton scattering at 100-200 MeV
\cite{seifert1993}.

Similarly, the spin rotation parameter $Q$ remains the most demanding observable to reproduce, as it isolates the real part of the spin-flip amplitude. Given that $Q$ is generally less constrained by current folding models, the moderate deviations observed are consistent with the known limitations of these
frameworks, rather than fundamental flaws in the underlying $\chi^\mathrm{EFT}$ potential.

\begin{figure}[htbp]
\centering
\includegraphics[width=0.99 \linewidth]{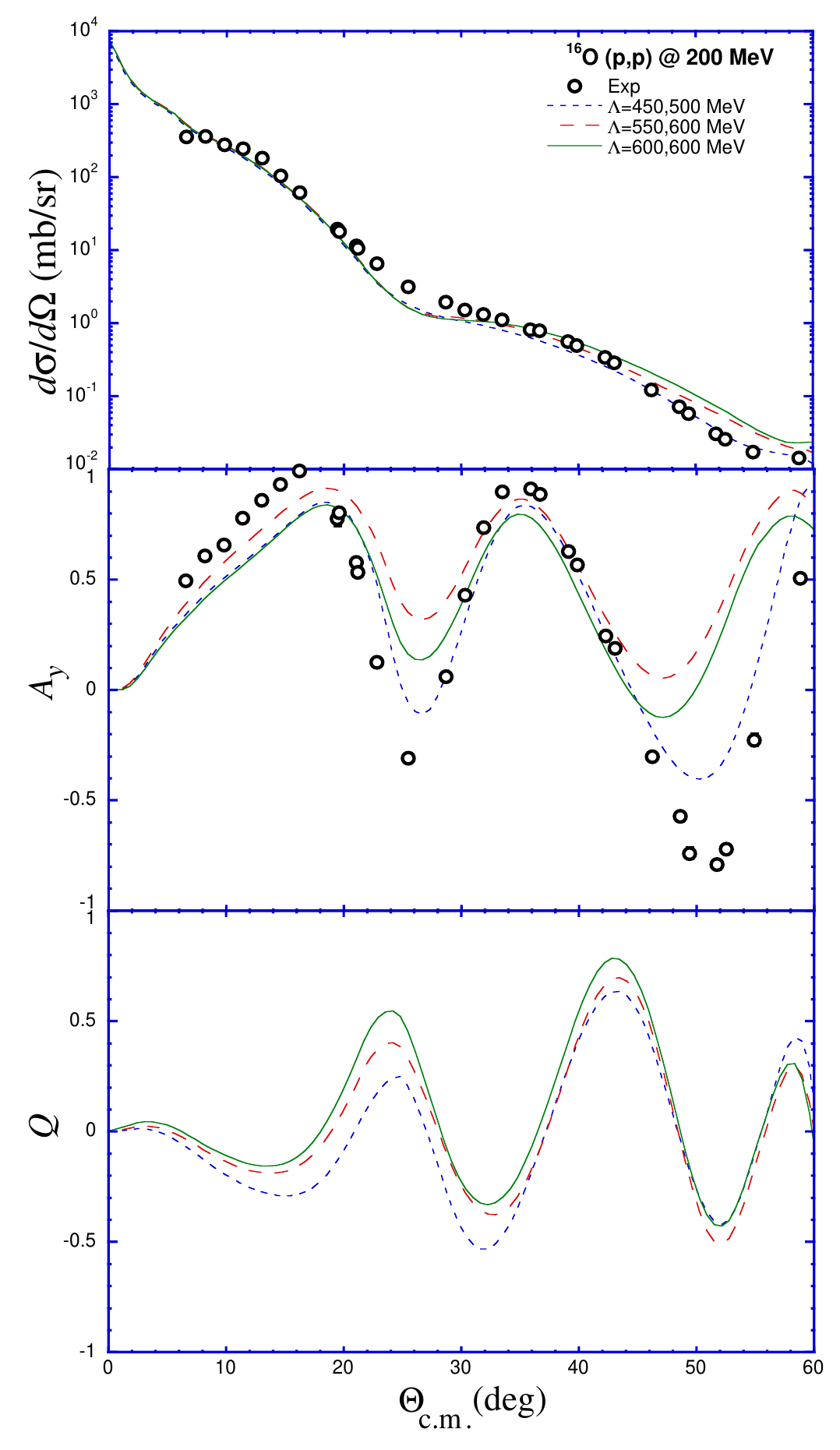}
\caption{Elastic scattering observables for 200 MeV protons on \ce{^{16}O}. Comparison of differential cross section ($d\sigma/d\Omega$), analyzing power ($A_y$), and spin rotation parameter ($Q$) calculated using
a single-folding optical potential based on N$^3$LO $\chi^\mathrm{EFT}$. Calculations utilize regulator cutoffs $(\Lambda,\tilde{\Lambda })$ of $(450, 500)$, $(550, 600)$, and $(600, 600)$ MeV.
Black circles are experimental data taken from the data tables of \cite{KellyNA} which compile the published results in Ref.\cite{seifert1993}; cross-section uncertainties are smaller than the symbols.}
\label{O16_200}
\end{figure}

Figure \ref{O16_200} is similar to figure \ref{O16_100} but for the elastic scattering of protons on \ce{^{16}O} at 200 MeV. The data exhibits an evolution in observable structure that complements the 100 MeV results,
providing a broader test of the N$^3$LO $\chi^\mathrm{EFT}$-based potential.

At 200 MeV the cross section follows the data over four orders of magnitude but is underestimated by a factor of about 2-3 between about $20^\circ$ and $33^\circ$. The model reproduces the nuclear surface and the gross geometry of the
nucleus. However, the diffraction minima are more pronounced and shifted toward smaller angles at 200 MeV, accompanied by an earlier onset of the theoretical "fan-out" in the cutoff
uncertainty band, reflecting the higher momentum transfers that probe shorter-range interactions.

Regarding spin observables, the oscillation pattern of the analyzing power $A_y$  and the positions of the diffraction minima are reproduced, but the depth of the minima is
underestimated, especially for the larger cutoffs. Nonetheless, this discrepancy is generally milder at 200 MeV, where the oscillatory patterns of $A_y$ and the spin rotation parameter $Q$ are more refined.
The increased oscillatory structure at 200 MeV, combined with the wider sensitivity to regulator cutoffs at higher energy, underscores the necessity of accounting for higher-order chiral corrections and medium-dependent
effects as the projectile energy increases. Collectively, these results demonstrate that while this model captures the essential physics of the scattering process, the growing theoretical uncertainty at higher energies
highlights the limits of the chiral expansion and the evolving role of many-body force contributions.

\begin{figure}[htbp]
\centering
\includegraphics[width=0.99 \linewidth]{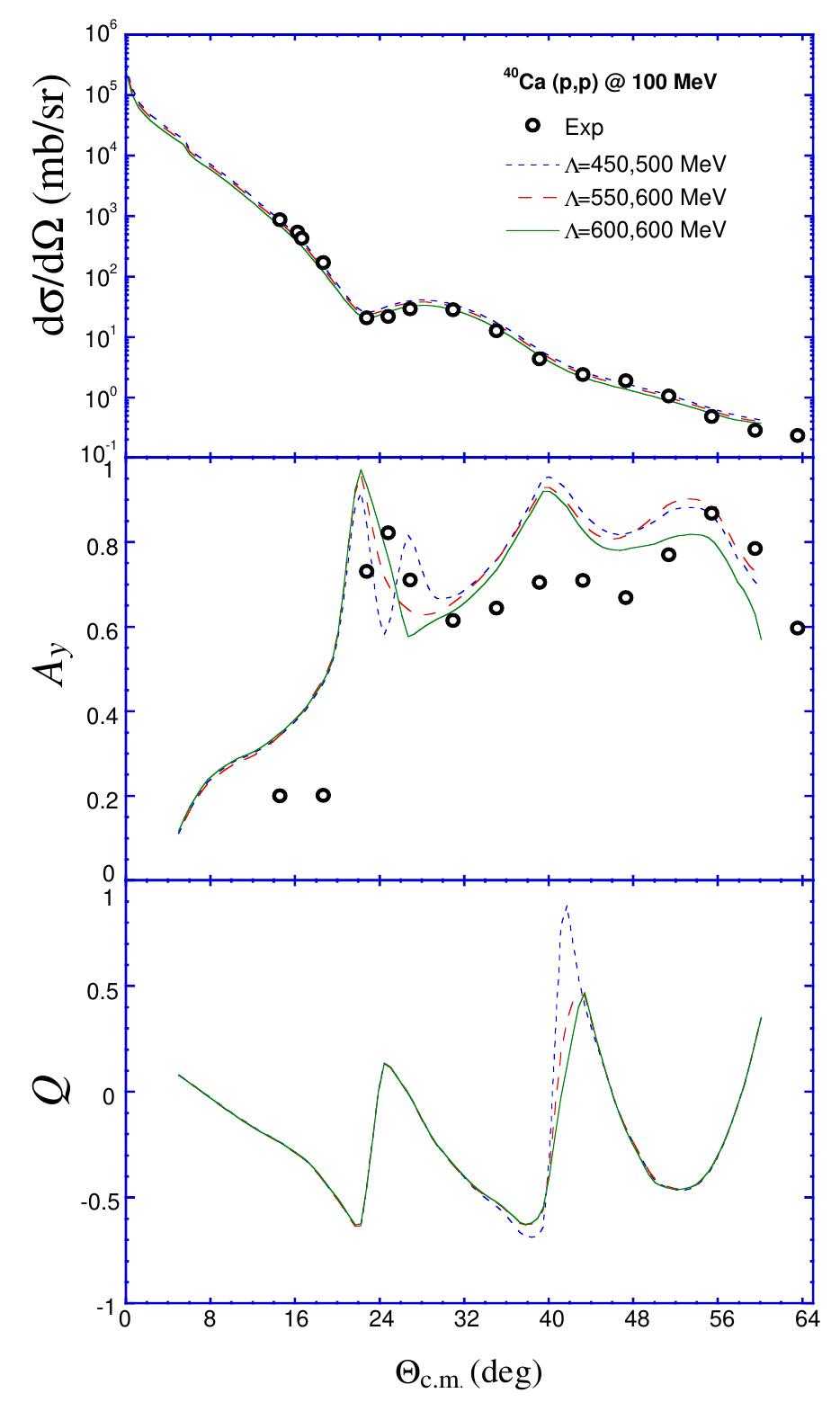}
\caption{Elastic scattering observables for 100 MeV protons on \ce{^{40}Ca}. Comparison of differential cross section ($d\sigma/d\Omega$), analyzing power ($A_y$), and spin rotation parameter ($Q$) calculated using
a single-folding optical potential based on N$^3$LO $\chi^\mathrm{EFT}$. Calculations utilize regulator cutoffs $(\Lambda,\tilde{\Lambda })$ of $(450, 500)$, $(550, 600)$, and $(600, 600)$ MeV.
Black circles are experimental data taken from the data tables of \cite{KellyNA} which compile the results of Ref.\cite{seifert1990} and the published results of in Ref.\cite{seifert1993};
cross-section and analyzing power uncertainties are smaller than the symbols.}
\label{Ca40_100}
\end{figure}

Figure \ref{Ca40_100} shows the observables for elastic scattering of protons by \ce{^{40}Ca} at 100 MeV. The differential cross section is reproduced reasonably well by all three regulators, with only mild cutoff
dependence; the curves separate appreciably only toward the large-angle minima, where the calculations are most sensitive to the interaction details. This indicates that the overall elastic yield is relatively stable
under regulator variation.

By contrast, the spin observables are markedly more sensitive to the cutoff choice -- more visible in the analyzing power near $22^\circ$--$27^\circ$ and in the spin rotation function near $40^\circ$--$42^\circ$.
The analyzing power is reproduced less well and develops more pronounced extrema for the softer (450, 500) MeV cutoff, whereas the harder cutoffs yield smoother angular behavior. The calculation overestimates the
data by about 0.1--0.2 at forward angles and between $30^\circ$ and $50^\circ$, and overshoots the sharp structure near $22^\circ$. These discrepancies are larger than the cutoff dependence, which suggests that they are not
due to the regulator choice but to the first-order truncation. This also follows from the fact that $A_y$ arises from the interference between spin-dependent and spin-independent amplitudes, so it responds strongly to shifts in their relative phase.
A softer regulator suppresses high-momentum components more aggressively and modifies the spin–orbit strength and off-shell behavior of the interaction, which displaces the positions and deepens the minima of $A_y$.

A similar trend appears in the spin rotation, where the soft cutoff produces a larger peak than the harder cutoffs. The spin rotation $Q$ measures the rotation of the polarization vector and is governed by the relative
phase between the non-spin-flip and spin-flip amplitudes. Because the soft regulator treats the spin–orbit and absorptive (imaginary) parts of the optical potential differently from the hard ones, it shifts this relative
phase and thereby enhances the extremum, producing the larger peak.

Overall, the figure shows that the differential cross section is comparatively robust under regulator variation, while the analyzing power and spin rotation remain sensitive to the short-range regularization.
Since residual cutoff dependence reflects the truncation uncertainty of the chiral expansion, this pattern indicates that the residual dependence is concentrated in the spin-dependent sector — consistent with
the spin–orbit and tensor components carrying a larger N$^3$LO truncation uncertainty than the central part. The spread among the spin-observable curves therefore provides a practical estimate of the chiral
truncation error and makes these observables particularly useful for testing the spin-dependent content of the optical $\chi^\mathrm{EFT}$ interaction.

\begin{figure}[htbp]
\centering
\includegraphics[width=0.99 \linewidth]{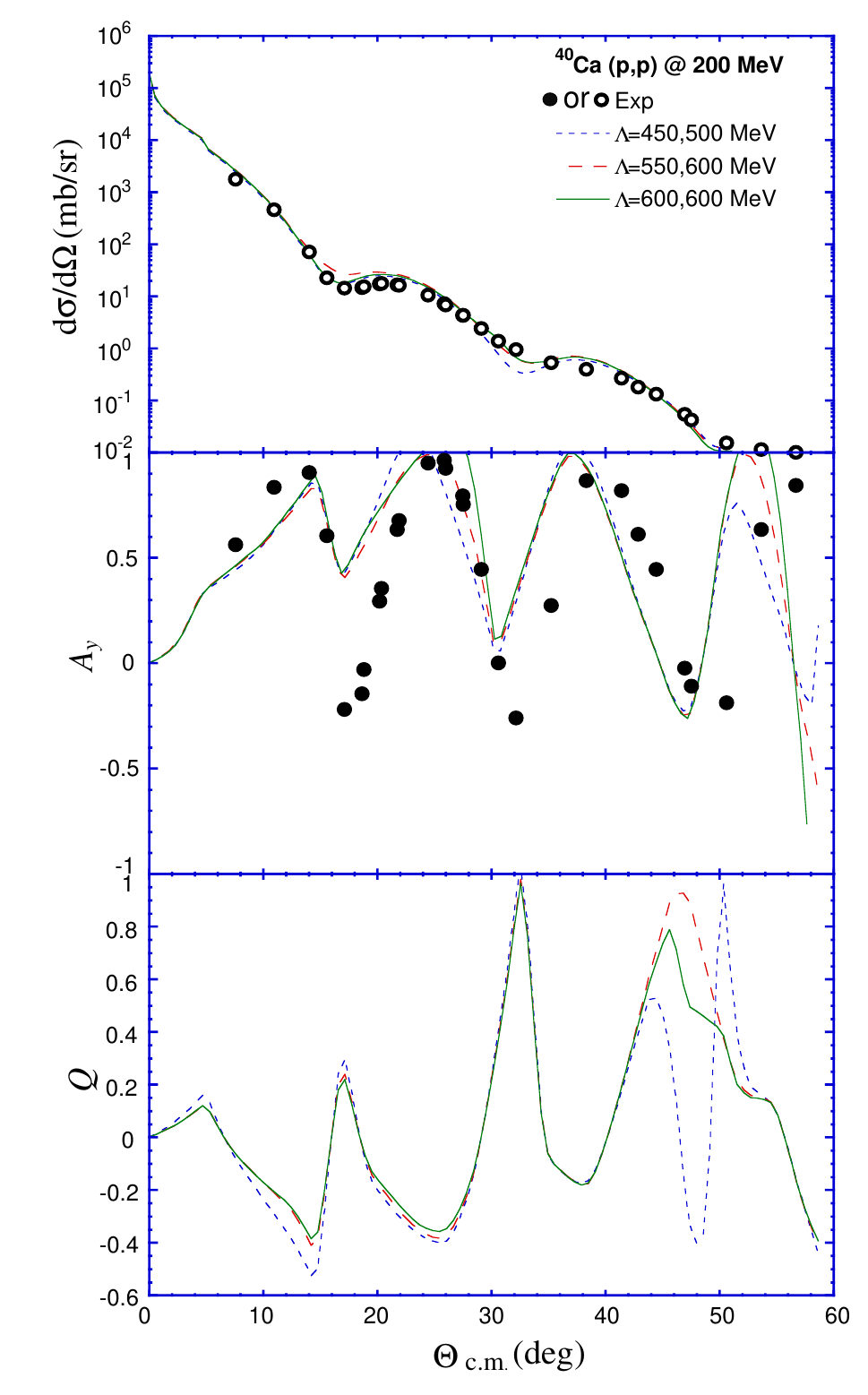}
\caption{Elastic scattering observables for 200 MeV protons on \ce{^{40}Ca}. Comparison of differential cross section ($d\sigma/d\Omega$), analyzing power ($A_y$), and spin rotation parameter ($Q$) calculated using
a single-folding optical potential based on N$^3$LO $\chi^\mathrm{EFT}$. Calculations utilize regulator cutoffs $(\Lambda,\tilde{\Lambda })$ of $(450, 500)$, $(550, 600)$, and $(600, 600)$ MeV.
Black circles are experimental data taken from the data tables of \cite{KellyNA} which compile the the published results in Ref.\cite{seifert1993};
cross-section and analyzing power uncertainties are smaller than the symbols.}
\label{Ca40_200}
\end{figure}

Figure \ref{Ca40_200} is analogous to figure \ref{Ca40_100}, but for 200 MeV protons. Compared with the 100 MeV case, the spin observables show a clearer increase in cutoff sensitivity, while the differential cross section remains broadly well reproduced by all three regulators. The cross-section agreement with the data is still good, but the angular structure is more demanding at 200 MeV, so the differences among regulator choices are easier to discern.

The analyzing power $A_y$ is more intricate than at 100 MeV. Although the three curves happen to converge in a few angular windows, the overall regulator spread is larger, and the experimental points carry more scatter. This reflects the higher projectile energy: more partial waves contribute and the phase accumulated across them is larger, so spin asymmetries become more sensitive to the details of the optical potential. The soft (450, 500) MeV cutoff still produces more pronounced extrema, but because the whole observable is already more structured at 200 MeV, the distinction is less "smooth versus structured" and more a matter of how the peaks and dips shift in angle.

The spin rotation is correspondingly more dramatic than at 100 MeV, with sharper extrema and stronger cutoff dependence. As an interference observable, $Q$ is highly responsive to changes in the phase and magnitude of the spin-dependent amplitudes; the larger phase accumulation at 200 MeV amplifies the regulator-induced differences into more visible shifts in peak positions and heights.

The overall picture across the two energies is consistent: 100 MeV gives the cleaner illustration of how the soft cutoff enhances the extrema of the spin observables, while 200 MeV shows that this sensitivity persists and is amplified by the more complex scattering dynamics -- and by the fact that 200 MeV probes the upper end of the energy range where the N$^3$LO expansion is well controlled. The regulator dependence stays modest in the cross section but grows in the spin observables, indicating that the residual cutoff dependence -- and hence the chiral truncation uncertainty -- is concentrated in the spin-dependent sector and increases with energy.

\section{Conclusion and outlook}
\label{sec:conclusion}
A fully microscopic, parameter-free description of $NN$ and $NA$ elastic scattering is developed using $\chi^\mathrm{EFT}$ theory at N$^3$LO. Taking the EGM potential with SFR cutoff $\tilde{\Lambda}$ as input, the LS equation is solved in momentum space -- convergence ensured by a Gaussian regulator cutoff $\Lambda$ -- to obtain the on-shell $NN$ $t$-matrix, projected it onto the Wolfenstein central amplitudes $B$ and spin-orbit amplitudes $C$, and folded it with relativistic mean-field densities to build a first-order optical potential, evaluating all observables at 100 and 200 MeV for three cutoff sets $(\Lambda,\tilde{\Lambda}) = (450, 500)$, $(550, 600)$, $(600, 600)$ MeV.

The framework performs best at 100 MeV, where the $pp$ and $pn$ amplitudes reproduce the empirical data in sign, magnitude, and angular shape, and the three cutoff curves cluster into a narrow band consistent with the expected truncation uncertainty. At 200 MeV the band broadens and the curves depart more from the data — most visibly in the spin-orbit and imaginary parts -- as the relative momentum approaches the chiral breakdown scale and the nearby pion-production threshold makes absorptive physics, least constrained in a first-order potential, increasingly important; the $pp$ channel degrades more than $pn$ owing to its smaller amplitudes, its restriction to isospin $T=1$, and Coulomb–nuclear interference. For \ce{p + ^{16}O} and \ce{p + ^{40}Ca} the differential cross sections agree well at forward and intermediate angles, where surface geometry dominates and regulator dependence is minimal, while the analyzing power $A_y$  and spin rotation $Q$ are markedly more cutoff-sensitive -- the residual dependence, and hence the chiral truncation uncertainty, is concentrated in the spin-dependent sector and grows with energy. The systematic underprediction of $A_y$ at its maxima and the larger deviations in $Q$ reflect well-documented limitations of first-order folding models rather than a defect of the EGM potential. Taken together, the EGM N$^3$LO potential with SFR provides a consistent and quantitatively reliable framework for $NN$ observables and $NA$ elastic scattering across 100–200 MeV, with its accuracy highest at 100 MeV and at forward-to-intermediate angles, and its growing truncation band and spin-sector discrepancies at 200 MeV marking the practical edge of a first-order, N$^3$LO description.

Future work should extend the $NN$ input to N$^4$LO to test whether the cutoff band contracts as the truncation-error estimates predict, and move beyond the first-order optical potential by adding second-order (medium and Pauli-blocking) terms together address the absorptive physics and the persistent $A_y$ and $Q$ discrepancies. Broadening the calculations above 200 MeV and below 100 MeV, applying the framework to heavier and exotic targets with consistent microscopic densities, and adopting a Bayesian truncation-error analysis to replace the qualitative cutoff spread with a rigorous uncertainty budget would further sharpen and quantify the predictive power of the approach.

\begin{acknowledgments}
The numerical calculations and computational simulations in this work were performed using the high-performance computing (HPC) cluster at the Department of Physics, College of Science, Qassim University.
I express my appreciation for the institutional infrastructure and technical support provided.
\end{acknowledgments}

\bibliography{references}

\end{document}